\documentclass[10pt,twoside,reqno]{amsart}
\usepackage{mathbbol,mathtools,slashed}

\usepackage{multirow}
\usepackage[bookmarksnumbered, plainpages]{hyperref}
\usepackage[rightcaption]{sidecap}
\usepackage{caption,xparse}
\usepackage{pgfplots}

\usepackage{float}

\usepackage[english]{babel}
\usepackage[latin1]{inputenc}
\usetikzlibrary{positioning,arrows,patterns}

\usepackage{graphicx}
\usepackage{epstopdf}
\usepackage{subfig}
\usepackage{color}
\usepackage{amsthm}
\usepackage{amsmath, amsfonts, amssymb}
\usepackage[figuresright]{rotating}
\begin{document}
   \title{Finite-Volume Non-extensive statistical in QCD phase}
   \author{\small M. A. A. AHMED$^{1,\dag}$
    H. ZAINUDDIN$^{2,\ddag}$ }
\thanks{\scriptsize emails: \\ $^\dag$mohammed\_h7@yahoo.com.\\ $^\ddag$ Hishamuddin.zainuddin@xmu.edu.my.}
\maketitle
\begin{center}
\address{
 $^{1}$ Physics Department, Faculty of Science, Taiz University Al-Turba branch, Taiz, Yemen.\\
 $^{2}$ Department of Physics, Faculty of Science, Universiti Putra Malaysia (UPM), Selangor, Malaysia.}
\end{center}
\begin{abstract}
Recently, some research has emerged that used the Tsallis statistics in Quantum Chromodynamics (QCD). In Tsallis statistics, the Tsallis parameter $\mathfrak{q}$ reflects long-range interactions, non-Markovian memory, and multifractal boundary effects, linking directly to temperature variance in fluctuating systems, which vanish in the Boltzmann-Gibbs limit. This is our motivation to investigate the thermodynamical variables of a hot and dense system. Within a non-extensive color QCD MIT-Bag Model and using the Tsallis distributions to describe the thermal response functions related to the equation of state.
The study focuses on the deconfinement phase transition from a hadronic gas to a quark-gluon plasma in order to analyze the behavior of thermodynamical quantities of the system, such as the energy density, the pressure and the interaction measure at finite volume under Tsallis statistics. Furthermore, the effect of the $\mathfrak{q}$-parameter on the phase transition point is investigated. We find that the transition point is significantly affected by the $\mathfrak{q}$-parameter in the framework of non-extensive statistics.
\end{abstract}
\markboth{\rightline {\sl M.A.A. Ahmed \& H. Zainuddin}}
        {\leftline{\sl Finite-Volume Non-extensive statistical in QCD phase}}

\section{\textbf{INTRODUCTION}}\
The study of strongly interacting matter is one of the most important topics in physics. This interaction is governed by Quantum Chromodynamics (QCD), which plays a crucial role in understanding the fundamental structure and dynamics of our universe \cite{Andersen2001,Andersen1992}. QCD is the theory of the strong nuclear force, explaining interactions between the elementary particels of hadrons, namely quarks and gluons, which are crucial for hadron structure, such as neutrons and protons \cite{Sarkar2009,Chodos1974}.

Central phenomena within QCD include color confinement, defined as the property of quarks being permanently bounded within hadrons, next to chiral symmetry breaking, which are both necessary for describing the nature of matter \cite{Cleymans1986,Robert1977}. Although lattice QCD presents a numerical approach based on first principles, phenomenological descriptions open relevant analytical points of view \cite{Banero2009}. Out of such descriptions, the MIT Bag Model is a remarkable framework, describing hadron confinement based on the hypothesis that quarks are confined within a "bag" which allows for their existence, protected from the perturbative QCD vacuum, due to a mismatch of energy density \cite{Chodos1974,DeGrand1975,Hasenfratz1977,Lavagno2009,Brito2011}.

The analysis of the QCD systems confined under volume confinement is of considerable significance \cite{Ejiri,Karsch2005,Ejiri2006,Ejiri2006a}. Under computational frameworks like lattice QCD, simulations must consequently be performed on a spacetime lattice that is discrete, finite; hence, a more profound analysis of volume effects at finite volumes is required for extrapolation of results suitable for physical contexts involving spacetimes of infinite volumes \cite{Ejiri2006a,Karsch2025,Gasser1987,Cheng2008}. Furthermore, volume considerations apply to several physical contexts, i.e., the early universe, compact astrophysical entities, and laboratory explorations on heavy-ion collisions (HIC), for which confinement at finite dimensions is a natural outcome of involved entities \cite{KAPUSTA,Spieles1998}. An investigation of how hadronic features, interactions of quark-gluon, and phase transformations, e.g., conversion of hadronic matters into Partonic Plasma (PP) or Quark Gluon Plasma (QGP) are affected by the finite boundaries is of utmost significance \cite{Cleymans1986,Robert1977,Zeldovitch1959,Zeldovitch1965,Ivanenko1969,Itoh1970}. Incorporating a finite volume within the realm of the MIT Bag Model permits the analytical analysis of exterior spatial limitations on the confinement of quarks, with subsequent consequences on subsequent hadronic features \cite{Huang2024,Klein2022}. These effects have the potential to greatly alter hadronic masses, interactions, and phase conversions, consequently giving details on the long-distance nature of QCD and interconnectivity between confinement and the loss of confinement (deconfinement) \cite{Binder1984,Binder2002,Binder1997,Djoudi2022,Djoudi2016,Imry1980}.

Namely, thermal QCD studies quark and gluon dynamics but struggles to predict hadron properties due to its nonperturbative character. The MIT bag model simplifies this by providing a coarse-grained model for hadron structure and confinement, using the bag constant $\mathfrak{B}$ to represent the energy cost of a QCD vacuum \cite{Cleymans1986,Robert1977}. Finite volume corrections and the partition function are of primary interest for thermodynamical property investigation and QCD matter phase transitions \cite{Rybczynski2014,Mitra2018}. The recent use of Tsallis Statistics (TS) is an intriguing outlook for systems that could have nonextensive attributes \cite{Tsallis1988,Tsallis2009}. Boltzmann-Gibbs (BG) standard statistical mechanics depends on the supposition of extensivity with short-range correlations. For extremely complex matter such as the QGP generated at high-energy collisions between heavy ions or within very highly correlated hadronic matter, the presence of long-range interactions, nonequilibrium, or fractal structures indicates the onset of departures from standard statistical conduct \cite{Wilk2000,Beck2000,Deppman2012,Marques2013,Marques2015,Azmi2015}. Furthermore, there is increasing evidence that TS can be considered as a more appropriate basis of a theoretical framework for describing complex systems whose properties cannot be exactly described by the BG statistical mechanics, such as the impact on Big Bang nucleosynthesis predictions of adopting a generalized distribution to describe the nucleons velocities \cite{Hou2017,Alberico2000}, and the estimated nuclear reaction rates and solar neutrino fluxes \cite{Lavagno2001}, and so on.
It is well known that BG statistics does not describe correctly the statistical behavior of particles with long-range interaction, non-Markovian memory and multifractal boundary conditions \cite{Tsallis1988,Lavagno2001}. When the speed of sound was studied in the hadronic medium by using TS for different $\mathfrak{q}$-values for a hadron resonance gas, as the TS allows for the exploration of systems that are away from thermal equilibrium \cite{Khuntia2016}.
Moreover, the chiral and deconfinement transitions were found to be crossover at finite temperature and without the chemical potential, regardless of the nonextensivity parameter $\mathfrak{q}$ \cite{Zhao2020}.
The evolution of the first-order energy density perturbation in hot, ideal, and non-extensive PP with a fluctuating environment was studied, and a non-extensive variant of the MIT bag model was considered to discover a breaking wave solution \cite{Bhattacharyya2020}.
While the incorporation of TS into the analysis of QCD and the MIT Bag Model within a finite volume has the potential for a finer determination of the thermodynamically relevant aspects of strongly interacting matter under extreme states \cite{Zhao2020,Lavagno2002,Bediaga2000,Conroy2010,Conroy2008,Teweldeberhan2004,Pennini1995,Shalaby2021,Sena2013}.
In thermal QCD, the partition function is the main quantity for studying thermodynamical quantities for same Thermal Response Functions (TRF), such as pressure, energy density, and entropy, which provide insights into possible phase transitions between the confined HG phase and the deconfined PP phase \cite{Mhamed2015,Mhamed2018,Mhamed2019,Mhamed2021}. The partition function encapsulates all relevant information about the system under investigation and has the following typical form:
\begin{equation}\label{par}
Z^{BG}(V;\beta,\mu)=\text{Tr}[e^{-\beta (\hat{\mathcal{H}}-\mu_k \hat{N}_k)}],
\end{equation}
where $\hat{\mathcal{H}}$ is the Hamiltonian of the system. Where $\beta$ represents the average inverse temperature, $k_B$ is Boltzmann's constant, $\mu_k$ is the chemical potential of the single-particle state $k$ and $\hat{N}_k$ is number operator for each single-particle state $k$ \cite{Ruelle1999}.

The model provides a simple but effective Equation of State (EoS) for strongly interacting matter, where the transition to the PP occurs once the thermal pressure inside the bag exceeds the vacuum pressure \cite{Chodos1974}. After that, the MIT bag model has been extensively applied to study the QCD phase transition at finite temperature $T$ and chemical potential $\mu$, the thermodynamics of PP, and astrophysical objects such as strange quark stars \cite{Gasser1987,Karsch2025}. Later extensions incorporated non-extensive statistical mechanics TS to better account for long-range correlations and deviations from equilibrium in high-energy collisions, as discussed in works by \cite{Lavagno2002,Bhattacharyya2013}.
The studies continue to refine the model to explore the QCD equation of state and critical behavior, making the MIT bag model a cornerstone for investigating confinement and deconfinement phenomena in QCD \cite{Zhao2020,Zhang2024}.
 In this work, we revisit the problem of phase transition from HG to PP with TS at finite volume. In this scenario, we address the connection between phase transition and the non-extensive TS for some TRF.

\section{\textbf{Methodology}}\
The BG is useful for studying many problems in physics. However, some of the phenomena required the general statistics called non-extensive TS which introduced by Tsallis \cite{Tsallis1988,Tsallis2009}. In TS there are two features the generalization of the internal energy as well generalization BG entropy. On another hand, there is $\mathfrak{q}$-Deformed algebraic or/and Quantum groups that allow studying the standard thermodynamics by using a defined set of commutation and noncommutative relations and Jackson derivative \cite{Lavagno2009,Brito2011,Jackson1909,Jackson1996}.
The theoretical framework of equilibrium statistical mechanics has found applications in very different areas of physics such as elementary particle physics within QCD, string field theories and physics of chaos including many-body physics, etc \cite{Zhang2025,Singh2024,Nelson2024}. It is well known that the partition function is a very important tool in statistical physics as Gibbs stated that the partition function contains all of the system's information \cite{Gibbs1902}.
\subsection{Boltzmann-Gibbs statistical mechanics}
The most important quantity in BG statistics is the entropy, using the Shannon's formula:
\begin{equation}\label{bge}
  \mathcal{S}^{BG}=-k_B\ \text{Tr}(\hat{\rho} \ln \hat{\rho}),
\end{equation}
with $\hat{\rho}$ density matrix trace corresponding to the total probability being equal to unity
\begin{equation}\label{p1}
  \text{Tr}\hat{\rho}=1.
\end{equation}
Gibbs, von Neumann, and Shannon further discussed this entropic form in detail \cite{Gibbs1902,Neumann1927,Shannon1948}.
Among the most remarkable fact in BG statistics is the additive property of entropy. For a system $\mathfrak{s}$ composed by any two independent subsystems $\mathfrak{s}_1$ and $\mathfrak{s}_2$ (i.e., satisfying $\hat{\rho}(\mathfrak{s}_1\cup \mathfrak{s}_2)=\hat{\rho}(\mathfrak{s}_1)\otimes\hat{\rho}(\mathfrak{s}_2)$, we can show that
\begin{equation}\label{BGentropy}
  \mathcal{S}^{BG}(\mathfrak{s}_1\cup \mathfrak{s}_2)=\mathcal{S}^{BG}(\mathfrak{s}_1)+\mathcal{S}^{BG}(\mathfrak{s}_2).
\end{equation}
As a result, the BG entropy is extensive in the sense that the sum of the entropies of two independent subsystems equals their total entropy. This property is lost when memory effects and long-range forces are present, and entropy, which is a measure of the information about the particle distribution in the accessible states, is no longer an extensive quantity \cite{Alberico2000}. In equilibrium statistical mechanics, the essential object is the statistical density matrix $\hat{\rho}$
\begin{equation}\label{den}
\hat{\rho}(V;\beta,\mu)=e^{-\beta (\hat{\mathcal{H}}-\mu_k \hat{N}_k)}.
\end{equation}
We defined the grand canonical partition function of the system in Eq.(\ref{par}) by defining the density matrix. The ensemble average of any thermal observable is calculated using the density matrix, which is represented by the operator $\hat{A}$ and defined by
\begin{equation}\label{tev}
\langle\beta| \hat{A}|\beta\rangle= \displaystyle \frac{\text{Tr}(\hat{A}\ \hat{\rho}(V;\beta,\mu))}{\text{Tr}[\hat{\rho}(V;\beta,\mu)]}=\text{Tr}(\hat{A}\ e^{-\beta (\hat{\mathcal{H}}-\mu_k \hat{N}_k)}).
\end{equation}
Now the grand canonical partition function for Bose-Einstein statistics is given by
\begin{equation}\label{zbe}
  Z^{(BE)}(V;\beta,\mu)=\prod_{k=1}\sum_{n_k=0}^{\infty}\left(e^{-\beta (\varepsilon_k-\mu)}\right)^{n_k}.
\end{equation}
As we see in bosons case, each occupation number is assigned a value $n_k=0,1,2,\cdots, N$ where occupation numbers satisfy
$$\sum_{k=1}^{\infty}n_k=N.$$
Also, the energy eigenvalues satisfy
$$\varepsilon=\sum_{k=1}^{\infty}n_k \varepsilon_k.$$
The sum in Eq.(\ref{zbe}) is a geometric series, and has the value
\begin{equation}\label{sumb}
\sum_{n_k=0}^{\infty}\left(e^{-\beta (\varepsilon_k-\mu)}\right)^{n_k}=\displaystyle \frac{1}{1-e^{-\beta (\varepsilon_k-\mu)}}.
\end{equation}
Hence, we can write Eq.(\ref{zbe}) as
\begin{equation}\label{fzbe}
  Z^{BE}(V;\beta,\mu)=\prod_{k=1}\displaystyle \frac{1}{1-e^{-\beta (\varepsilon_k-\mu)}}.
\end{equation}
On the other hand, for fermions the occupation numbers are restricted to the values $n_k=0,1$ because of Pauli's exclusion principle. For Fermi-Dirac statistics, one obtains,
 \begin{equation}\label{fzfd}
 Z^{FD}(V;\beta,\mu)  =\prod_{k=1}\left(1+e^{-\beta (\varepsilon_k-\mu)}\right ),
\end{equation}
Now, we can write both partition functions for the bosons and fermions in Eqs.(\ref{fzbe}) and (\ref{fzfd}) as
\begin{equation}\label{partition2}
  Z^{(BE,FD)}(V;\beta,\mu)=\prod_{k=1}^{n}\left(1+\alpha e^{-\beta(\varepsilon-\mu)}\right)^{\alpha },
\end{equation}
where $\alpha$ is parameter refer to bosons for $\alpha=-1$ or fermions for $\alpha=1$. When the volume of the system is large, so the limit of volume can be recovered with the normal replacement
\begin{equation}\label{partition20}
\sum_{k}\longrightarrow V\int \displaystyle \frac{d^3p}{(2\pi)^3}.
\end{equation}
Recalling the mode partition from Eq.(\ref{partition2}), hence, we use the transformation in Eq.(\ref{partition20}) to write
\begin{equation}\label{partition3}
  \ln Z^{\text{(BE,FD)}}(V;\beta,\mu)=V \int \displaystyle \frac{d^3p}{(2\pi)^3} \ln\left\{\left(1+\alpha e^{-\beta(\varepsilon-\mu)}\right)^{\alpha}\right\}.
\end{equation}
 The mean number of the occupation number particle in BG statistics is given by
\begin{equation}\label{np}
  \langle \hat{n}_k^{\text{(BE,FD)}}\rangle=n^{\text{(BE,FD)}}(V;\beta,\mu)=\displaystyle \frac{1}{e^{\beta (\varepsilon_k-\mu)}+\alpha}.
\end{equation}
There is no restriction i.e. $\mu\in[-\infty,\infty]$.
\subsection{Tsallis statistical mechanics}
There are new researchers who use the non-extensive TS as a validation test of the BG statistics, which is a specific instance of the generalized statistical mechanics, when the Tsallis parameter $\mathfrak{q}$ is equal to unity. The entropic parameter $\mathfrak{q}$ describes the Tsallis distribution \cite{Wilk2000,Ruelle1999}. This parameter is connected to the relative temperature variance in a system of fluctuating temperature zones. These fluctuations vanish for BG distribution \cite{Wilk2000}.
As mentioned before, TS is the generalization of BG statistics \cite{Tsallis1988,Tsallis2009}
\begin{equation}\label{tge}
  \mathcal{S}_{\mathfrak{q}}=k_B\ \displaystyle \frac{\text{Tr}(\hat{\rho}-\hat{\rho}^{\mathfrak{q}})}{\mathfrak{q}-1} \ \ \ (\mathfrak{q}\in\mathbb{R}).
\end{equation}
The entropy $\mathcal{S}^{\mathfrak{q}}$ can be conveniently rewritten in the following alternative form:
\begin{equation}\label{tge1}
  \mathcal{S}_{\mathfrak{q}}=-k_B \ \text{Tr}(\hat{\rho}\ln_{\mathfrak{q}}\hat{\rho}),
\end{equation}
where the $\mathfrak{q}$-logarithmic function is defined as follows:
$$\ln_{\mathfrak{q}}\hat{\rho}=\displaystyle \frac{\hat{\rho}^{\mathfrak{q}-1}-1}{\mathfrak{q}-1}.$$
In the limit $\mathfrak{q}\longrightarrow 1$, the usual natural logarithm is restored. It is called the generalized logarithm because it is the inverse function of the generalised exponent:
\begin{equation}\label{apr}
e_{\mathfrak{q}}^{\hat{\rho}}=\left(1+\hat{\rho}(\mathfrak{q}-1)^{ \frac{1}{\mathfrak{q}-1 }}\right).
\end{equation}
Tsallis entropy of a system $\mathfrak{s}$ when it is composed by any two independent subsystems $\mathfrak{s}_1$ and $\mathfrak{s}_2$ (i.e., satisfying $\hat{\rho}_{\mathfrak{q}}(\mathfrak{s}_1\cup \mathfrak{s}_2)=\hat{\rho}_{\mathfrak{q}}(\mathfrak{s}_1)\otimes\hat{\rho}_{\mathfrak{q}}( \mathfrak{s}_2)$, it satisfies the following relation
\begin{equation}\label{TGentropy}
  \mathcal{S}_{\mathfrak{q}}(\mathfrak{s}_1\cup \mathfrak{s}_2)=\mathcal{S}_{\mathfrak{q}}(\mathfrak{s}_1)+\mathcal{S}_{\mathfrak{q}}(\mathfrak{s}_2)+\displaystyle (1-\mathfrak{q})\mathcal{S}_{\mathfrak{q}}(\mathfrak{s}_1)\mathcal{S}_{\mathfrak{q}}(\mathfrak{s}_2).
\end{equation}
Therefore, Tsallis entropy is generically non-extensive (non-additive).
The aforementioned assumptions are tested in high-energy nuclear collisions, and experimental signals are observed that can be interpreted as a result of the presence of a non-extensive regime \cite{Alberico2000}.
The analogous density matrix to Eq.(\ref{den}) for TS is given by
\begin{equation}\label{dentg}
\hat{\rho}(V;\beta,\mu)=\displaystyle \frac{\left(1+(\mathfrak{q}-1)\beta \hat{\mathcal{H}}\right)^{ \frac{1}{1-\mathfrak{q}}}}{Z_{\mathfrak{q}}(V;\beta,\mu)}.
\end{equation}
The Tsallis density matrix is used to compute the ensemble average of any thermal observable, represented by the operator $\hat{A}$, then, defined as
\begin{equation}\label{ttev}
\langle\beta| \hat{A}|\beta\rangle_{\mathfrak{q}}=\displaystyle \frac{\text{Tr}(\hat{A}\ \hat{\rho}(V;\beta,\mu))}{\text{Tr}[\hat{\rho}(V;\beta,\mu)]},
\end{equation}
while the $\mathfrak{q}$-partition function of system is given by
\begin{equation}\label{part}
Z_{\mathfrak{q}}(V;\beta,\mu)=\text{Tr}'[1+(\mathfrak{q}-1)\beta \hat{\mathcal{H}}]^{ \frac{1}{1-\mathfrak{q}}}.
\end{equation}
The stated trace imposes the Tsallis cut-off condition, which means that it is taken over the energy eigenvalues satisfying the constraint $1+(\mathfrak{q}-1)\beta (\varepsilon-\mu)\geq0$.
When the parameter $\mathfrak{q}<1$, the FD and BE distributions have a natural high-energy cut-off: $\varepsilon \leq1/(\mathfrak{q}-1)\beta +\mu$, implying that the energy tail is depleted; when $\mathfrak{q}>1$, the cut-off is absent, implying that the energy tail of the particle distribution is enhanced for fermions and bosons \cite{Alberico2000}.
If we follow the same techniques which is done for the BG partition function in Eq.(\ref{partition2}), the Tsallis partition function will be written as
\begin{equation}\label{tpartition3}
  Z^{\text{(BE,FD)}}_{\mathfrak{q}}(V;\beta,\mu)=\prod_{k=1}^{n}\left(1+\alpha e_{\mathfrak{q}}^{-\beta(\varepsilon-\mu)}\right)^{\alpha}.
\end{equation}
The mean number of the occupation number in TS is given by
\begin{equation}\label{tgparticles}
  \langle\hat{n}_{\mathfrak{q}}^{\text{(BE,FD)}}\rangle=n_{\mathfrak{q}}^{\text{(BE,FD)}}(V;\beta,\mu)=\displaystyle \frac{1}{(1-(\mathfrak{q}-1)\beta (\varepsilon-\mu))^{\delta}+\alpha}.
\end{equation}
We have to mention that the parameter $\delta=\displaystyle \frac{\mathfrak{q}}{\mathfrak{q}-1}$ can be got from the statistical mechanical formulations suggested by Tsallis in \cite{Bhattacharyya2020,Tsallis1998,Hasegawa2009}. In contrast, another form of the parameter in $\delta=\displaystyle \frac{1}{\mathfrak{q}-1}$ will have a slightly different form of the Tsallis quantum distributions \cite{Mitra2018,Zhao2020,Conroy2010,Conroy2008,Teweldeberhan2004,Pennini1995,Shalaby2021,Demirhan1993}. Now, we will show how TS go to BG statistics by taking $\mathfrak{q}$ tends to 1:
\begin{equation}\label{tgtobgent}
  \lim_{\mathfrak{q}\longrightarrow 1} \mathcal{S}_{\mathfrak{q}}(\mathfrak{s})=\mathcal{S}^{BG}(\mathfrak{s}),
\end{equation}
and, for the partition function, we obtain
\begin{equation}\label{tgtobgent0}
  \lim_{\mathfrak{q}\longrightarrow 1}Z_{\mathfrak{q}}(V;\beta,\mu)=Z^{BG}(V;\beta,\mu),
\end{equation}
Similarly, for the distribution of the particles i.e.
\begin{equation}\label{tgtobgent1}
  \lim_{\mathfrak{q}\longrightarrow 1}n_{\mathfrak{q}}(V;\beta,\mu)=n^{BG}(V;\beta,\mu).
\end{equation}

\section{\textbf{Tsallis Partition Function of the system}}\
The Hadronic Gas is considered as an ideal relativistic gas of massless pions, while the QGP phase is merely a free partonic system. We employ a basic Phase Coexistence Model (PCM) from \cite{Spieles1998}, for this purpose, in which the mixed phase system has a finite size $V=V_{HG}+V_{PP}$. The fraction of volume occupied by the HG phase (specified by the parameter $h$) is $V_{HG}=hV$, while the remaining volume is $V_{PP}=(1-h)V$, which contains the PP phase. We can simply express the total partition function of the system as a product of two partition functions if we ignore reciprocal interactions between the various phases:
\begin{equation}\label{totalpartition}
Z_{\mathfrak{q}}^{\text{SYS}}(h,T,V,\mu)=Z_{\mathfrak{q}}^{PP}(h,V;T,\mu)Z_{\mathfrak{q}}^{HG}(h,V;T,\mu).
\end{equation}
In the continuous case, we can write Eq.(\ref{tpartition3}), so the resulting expression for the logarithm of the Tsallis grand canonical partition function of particles and anti-particles with massless and chemical potential $\mu$ and degeneracy factor $g$ in the large volume limit of free gas is obtained as,
\begin{equation}\label{cqranalpartition}
  \ln Z_{\mathfrak{q}}^{\text{(BE,FD)}}(V;\beta,\mu)=\frac{g V}{6\pi^2}\int_{0}^{\infty}\left[\frac{p^3}{\left(1+(\mathfrak{q}-1)\beta(\varepsilon+\mu)\right)^{\frac{\mathfrak{q}}{\mathfrak{q}-1}}+\alpha}
  +\frac{p^3}{\left(1+(\mathfrak{q}-1)\beta(\varepsilon-\mu)\right)^{\frac{\mathfrak{q}}{\mathfrak{q}-1}}+\alpha}\right]dp.
\end{equation}
Now we will go to compute the Tsallis partition function for fermion and boson:
\begin{itemize}
  \item the Tsallis partition function of the fermi and anti-fermi gas,
  \begin{equation}\label{fqranalpartition}
  \ln Z_{\mathfrak{q}}^{F\bar{F}}(V;\beta,\mu)=\frac{g_{F\bar{F}} V}{6\pi^2}\sum_{s=1}^{\infty} (-1)^{s+1}\int_{0}^{\infty}\left[\frac{p^3}{\left(1+(\mathfrak{q}-1)\beta(\varepsilon\pm\mu)\right)^{\frac{s\mathfrak{q}}{\mathfrak{q}-1}}}\right]dp
\end{equation}
  \item the Tsallis partition function of the boson and anti-boson gas,
  \begin{equation}\label{bqranalpartition}
  \ln Z_{\mathfrak{q}}^{B\bar{B}}(V;\beta,\mu)=\frac{g_{B\bar{B}} V}{6\pi^2}\sum_{s=1}^{\infty} \int_{0}^{\infty}\left[\frac{p^3}{\left(1+(\mathfrak{q}-1)\beta(\varepsilon\pm\mu)\right)^{\frac{s\mathfrak{q}}{\mathfrak{q}-1}}}\right]dp
\end{equation}
\end{itemize}
In this work, we study the TS without chemical potential. For the HG and PP, the Tsallis partition function is given by,
\begin{equation}
\ln Z_{\mathfrak{q}}^{\text{(HG,PP)}}(V_{\text{(HG,PP)}};\beta)=\frac{g_{\text{(HG,PP)}}}{6\pi^2}V_{\text{(HG,PP)}}\beta I_{\mathfrak{q}}^{(B,F)}(\beta),
\end{equation}
and we can write the Tsallis partition function with order parameter $h$ as:
\begin{itemize}
  \item the Tsallis partition function of HG (Pion),
  \begin{equation}\label{hgpartition}
Z_{\mathfrak{q}}^{HG}(h,V;T)=Exp^{\left[VT^3h \left(\frac{T_{\mathfrak{q}}^B}{2\pi^2}\right)\right]}.
\end{equation}
\item the Tsallis partition function of PP,
\begin{equation}\label{qqpartition}
Z_{\mathfrak{q}}^{PP}(h,V;T)=Z_{\mathfrak{q}}^{\text{VAC}}(h,V;T)Z_{\mathfrak{q}}^{Q}(h,V;T)Z_{\mathfrak{q}}^{G}(h,V;T).
\end{equation}
  \item the Tsallis partition function of Gluons,
\begin{equation}\label{ggpartition}
Z_{\mathfrak{q}}^{G}(h,V;T)=Exp^{\left[VT^3(1-h) \left(\frac{8T_{\mathfrak{q}}^B}{3\pi^2}\right)\right]}.
\end{equation}
  \item the Tsallis partition function of Quarks,
\begin{equation}\label{qqpartition}
Z_{\mathfrak{q}}^{Q}(h,V;T)=Exp^{\left[VT^3(1-h) \left(\frac{2T_{\mathfrak{q}}^F}{\pi^2}\right)\right]}.
\end{equation}
  \item the Tsallis partition function of vacuum,
\begin{equation}\label{vacpartition}
Z_{\mathfrak{q}}^{\text{VAC}}(h,V;T)=Exp^{-\left[VT^3(1-h)\left(\frac{\mathfrak{B}}{T^4}\right)\right]},
\end{equation}
\end{itemize}
The difference between the real vacuum and the perturbative vacuum due to the color confinement meaning that the constant $\left(\mathfrak{B}\right)$ of the bag model represents the pressure on the surface of the bag in order to balance the outward pressure exerted by the partons moving inside the bag.
A non-variable Bag constant $\mathfrak{B}$ is not sufficient because the non-abelian nature of QCD leads to a complicated non-perturbative structure of QCD vacuum. In this work, we have always taken into account a constant value equal to 145, as this value is known to lie within the stability window that satisfies the Bodmer-Witten conjecture \cite{Torres2013}.

For a QGP consisting of gluons and two flavors of massless quarks (up, down) without a chemical potential $\mu$ in the bag model, by inserting Eqs.(\ref{hgpartition}), (\ref{ggpartition}), (\ref{qqpartition}) and (\ref{vacpartition})) into Eq.(\ref{totalpartition}), the total Tsallis partition function of the whole system will be written as:
\begin{equation}\label{totalpartition1}
Z_{\mathfrak{q}}^{\text{SYS}}(h,V;T)=Exp^{VT^3\left[\left\{-\frac{1}{\pi^2}\left(\frac{13}{6}T_{\mathfrak{q}}^B+2T_{\mathfrak{q}}^F\right)+\frac{\mathfrak{B}}{T^4}\right\}h+ \left\{\frac{2}{\pi^2}\left(\frac{4}{3}T_{\mathfrak{q}}^B+2T_{\mathfrak{q}}^F\right)-\frac{\mathfrak{B}}{T^4}\right\}\right]}.
\end{equation}
The result in Eq.(\ref{totalpartition1}) was obtained when we separated the parts biased the order parameter $h$ together and we repeated the process for other parts. This result will be the same as that gotten in the BG when $\mathfrak{q}\longrightarrow1$, so the limit of Eq.(\ref{totalpartition1}) at this value reads \cite{Mhamed2015,Mhamed2018,Mhamed2019},
\begin{equation}\label{BGpartition}
\lim_{\mathfrak{q}\to1} Z_{\mathfrak{q}}^{\text{SYS}}(h,V;T)=Exp^{VT^3\left[\left(\frac{37}{90}\pi^2-\frac{\mathfrak{B}}{T^4}\right)+\left(\frac{\mathfrak{B}}{T^4}-\frac{17}{45}\pi^2\right)h\right]}.
\end{equation}
Now, we define the Tsallis Hadronic Probability Density Function (THPDF) is given by
\begin{equation}\label{hpdf}
p_{\mathfrak{q}}(h,V;T)=\frac{Z_{\mathfrak{q}}^{\text{SYS}}(h,V;T)}{\int\limits_{0}^{1}Z^{\text{SYS}}_{\mathfrak{q}}(h,V;T)dh}.
\end{equation}
Because our THPDF is directly connected to the Tsallis partition function of the system, it's assumed that it contains all of the information about QCD matter and its deconfinement phase transition. The mean value of any TRF may therefore be calculated as follows: $Q_{\mathfrak{q}}(h,V;T)$, describing the system in this state $h$ as follows:
\begin{equation}
Q_{\mathfrak{q}}(V;T) =\int\limits_{0}^{1}Q\left(
h,V;T\right) p_{\mathfrak{q}}(h,V;T) dh.
\label{meanh}
\end{equation}

\section{\textbf{Finite Size QCD Thermodynamics and Equation Of State}}
In statistical mechanics and critical phenomena, infinite systems exhibit true singularities (e.g., divergence of heat capacity or susceptibility). However, in finite-size systems, these singularities are rounded and shifted, producing finite peaks instead of true divergences. This behavior is captured in finite-size scaling theory, which relates the system size $L=V^{1/d}$ to deviations from critical behavior in the thermodynamical limit. These singularities are converted into finite peaks with well-defined amplitudes and widths, in addition to the scaling of the thermodynamical quantities, and will be investigated as far as for TRF at finite size \cite{Mhamed2015,Mhamed2021,Fisher1972,Binder1981}.
\subsection{The Importance of the EoS in the Hydrodynamical Description}
Remember that the system under study and produced in URHIC expands as soon as it is created. It includes many particles in a small finite size on both sides of the deconfinement phase transition (partons in the PP and pions in the HG). If these particles interact strongly enough, the system may achieve local thermodynamical equilibrium. The relativistic hydrodynamics may readily represent the subsequent evolution of the PP and HG, if it can be locally maintained during the later growth. Hydrodynamics is a macroscopic method that characterises the system using macroscopic quantities e.g. local energy density $\varepsilon_{\mathfrak{q}}(V;T)$, pressure $p_{\mathfrak{q}}(V;T)$, the dynamical number of degrees of freedom and entropy density $s_{\mathfrak{q}}(V;T)$.
It needs familiarity with the EoS, which establishes a relationship between pressure, energy, and entropy, but not in-depth details of microscopic dynamics. This is what we hope to do by examining these TRF and their correlations.
In general, we require an EoS to represent the hydrodynamic expansion of the system of particles generated in these collisions in phenomenological methods to URHIC physics. The kind of EoS has a major influence on the space-time pattern of a collision, and the phase transition causes the collision duration to rise. The detailed study of the temperature dependence of some TRF such as pressure $p_{\mathfrak{q}}(V;T)$, and energy density $\varepsilon_{\mathfrak{q}}(V;T)$, as well as the pressure and sound velocity as function in energy $p_{\mathfrak{q}}(\varepsilon)$, is required to know more a detailed about these thermal functions in experimental and theoretical studies.
\subsection{Thermal Response Functions}
There are many TRF, so we will choose some of them to evaluate the phase transition at the finite size. However, we consider the selected functions in this study within TS.
The mean value of the hadronic volume fraction $H_{\mathfrak{q}}(V;T)$, which is used as the order parameter for the phase transition studied in this work, was the first quantity of considering in our research. These quantities can be represented according to Eqs.(\ref{hpdf}) and (\ref{meanh}) by,
\begin{equation}\label{meanH}
H_{\mathfrak{q}}(V;T)=\langle h(V;T)\rangle=\int\limits_{0}^{1}h\ p_{\mathfrak{q}}\left( h,V;T\right) dh.
\end{equation}
As a result, Eq.(\ref{meanH}) reads,
\begin{equation}\label{meanH1}
H_{\mathfrak{q}}(V;T)=\frac{1}{VT^3\left(\frac{1}{\pi^2}\left[\frac{13}{6}T_{\mathfrak{q}}^B+2T_{\mathfrak{q}}^F\right]-\frac{\mathfrak{B}}{T^4}\right)}
+\frac{1}{1-Exp^{VT^3\left(\frac{1}{\pi^2}\left[\frac{13}{6}T_{\mathfrak{q}}^B+2T_{\mathfrak{q}}^F\right]-\frac{\mathfrak{B}}{T^4}\right)}}.
\end{equation}
\begin{figure}[!h]
  \centering
 \includegraphics[width=250pt]{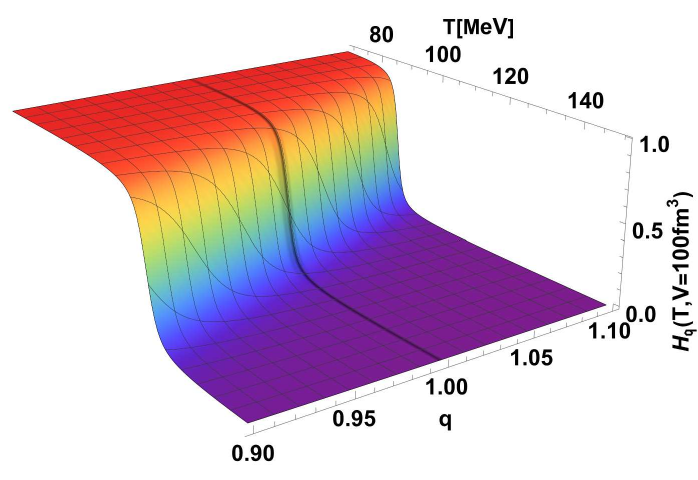}
  \caption{$H_{\mathfrak{q}}(V=100fm^3;T)$ is the order parameter of the system as a function of $(T,\mathfrak{q})$.}\label{figh}
\end{figure}
We will see that the most important physical quantities can be represented via these fundamental quantity. At low temperatures ($T\thickapprox 0.08-0.10 GeV$), the order parameter saturates near unity, indicating that the system remains entirely in the HG phase.
As the temperature increases toward the critical region ($T\thickapprox 0.12 GeV$), the order parameter decreases sharply, signaling the onset of the deconfinement transition, where hadrons dissolve into quarks and gluons to form the QGP.
Beyond the critical temperature $T_0$, the order parameter becomes small, reflecting that the system predominantly resides in the QGP phase at higher temperatures.
The order parameter thus serves as a quantitative measure of the degree of deconfinement-approaching unity in the HG phase, vanishing in the QGP phase, and taking intermediate values in the mixed transition region.
\begin{figure}[h]
  \centering
 \includegraphics[width=400pt]{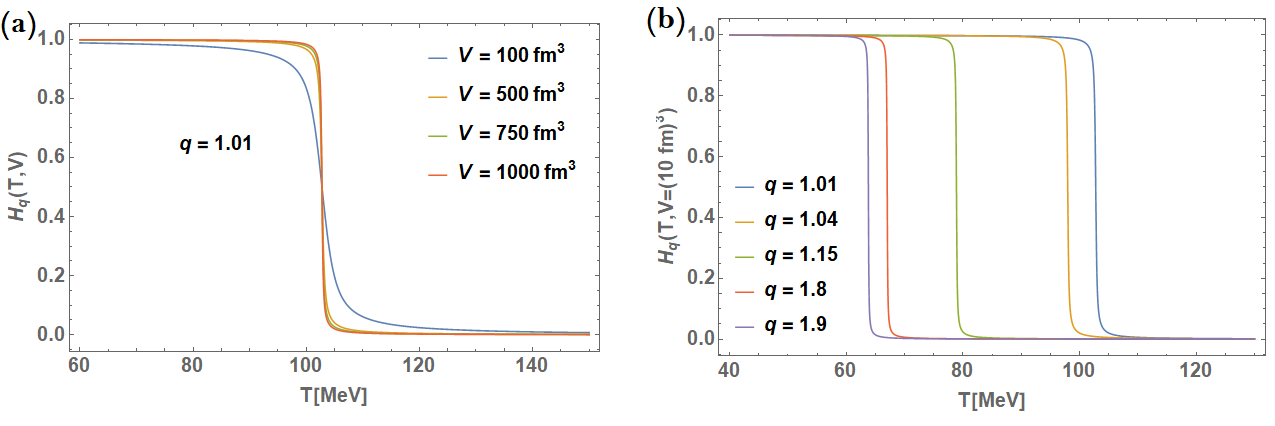}
  \caption{(a) The order parameter versus temperature $T$, when $\mathfrak{q}=1.01$ with set volumes $V=(0.1,0.25,0.5,1)(10fm)^3$.(b) The order parameter versus temperature $T$, when $V=(10fm)^3$ with set Tsallis parameter $\mathfrak{q}=(1.01,1.04,1.15,1.8,1.9)$.}\label{figh1}
\end{figure}
The Fig.(a-\ref{figh1}) shows us that there is no shifting effect of the transition point for different volumes with the same value of $\mathfrak{q}$, while the shifting effect of the transition point presents for different values of $\mathfrak{q}$ for the same size as we see that in Fig.(b-\ref{figh1}). The Fig.(b-\ref{figh1}) shows us the behaviour of the order parameter $H_{\mathfrak{q}} (V;T)$, we can consider that the transition point as the point in which we have equal probability of hadronic phase and PP phase: $H_{\mathfrak{q}} (V;T)=1-H_{\mathfrak{q}} (V;T)$. This indicates that $H_{\mathfrak{q}} (T_0(V))=1/2$ determines the value of the order parameter for both phases. Moreover, we know that the order parameter has a finite discontinuity in thermodynamical limit, which may be readily represented by a step function: $\lim_{\substack{V \to \infty \\ \mathfrak{q} \to 1}}H_{\mathfrak{q}}(V;T)=1-\Theta(T-T_0(\infty))$.
\begin{table}[h]
	
\begin{tabular}{|l|c|}
  \hline
  Tsallis parameter $\mathfrak{q}$ &  \begin{tabular}{l}
                                    The temperature of the transition point \\
                                        $T_0(V=100fm^3)$ [MeV]
                                      \end{tabular}
   \\
  \hline
$\mathfrak{q}=1.00$&  $104.2954\pm 0.00001$ \\
   \hline
$\mathfrak{q}=1.01$&  $102.6356\pm 0.00001$\\
   \hline
$\mathfrak{q}=1.04$&  $97.8636\pm 0.00001$\\
   \hline
$\mathfrak{q}=1.15$&  $78.7756\pm 0.00001$\\
   \hline
$\mathfrak{q}=1.80$&  $66.9490\pm 0.00001$\\
   \hline
$\mathfrak{q}=1.93$&  $60.1026\pm 0.00001$\\
   \hline
\end{tabular}
\caption{$T_0(V)$ is the temperature of the transition point for different Tsallis parameter values with volumes.}\label{8tab}
\end{table}
We can observe from the table.(\ref{8tab}) the transition point for different of $\mathfrak{q}$ values in volume $100 fm^3$, these points will be the same for various volumes, because we consider the colour condition of PP, while the shifting of the transition point happened for different volumes when we considered the colourless condition of QGP \cite{Mhamed2015,Mhamed2018,Mhamed2019,Mhamed2021}.

\begin{figure}[h]
\centerline{\includegraphics[width=250pt]{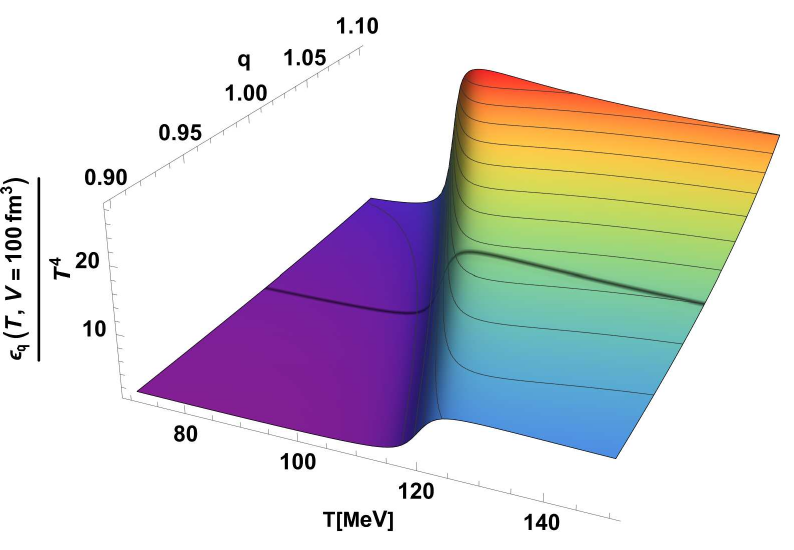}}
\caption{3-Dim normalized energy density $\frac{\varepsilon_{\mathfrak{q}}(T)}{T^{4}}$ of the system as a function of $(T)$, when volume $V=100fm^3$. \label{fig:FIG2}}
\end{figure}

Now, we introduce the second quantity of considering was the energy density $\varepsilon_{\mathfrak{q}}(V;T)$, which mean value was also computed in the same manner and was confirmed to be related to $H_{\mathfrak{q}}(V;T)$ by the relationship
\begin{equation}\label{EDensity}
  \varepsilon_{\mathfrak{q}}(V;T) =\frac{T^{2}}{V} \left<\left( \frac{\partial Ln Z_{\mathfrak{q}}(h,V;T)
}{\partial T}\right) \right>,
\end{equation}
after we use Eq.(\ref{EDensity}) the density energy will be given by
\begin{equation}\label{EDensity1}
  \frac{\varepsilon_{\mathfrak{q}}(V;T)}{T^4}=\left(\left(\frac{6}{\pi^2}\left[\frac{4}{3}T_{\mathfrak{q}}^B+2T_{\mathfrak{q}}^F\right]+\frac{\mathfrak{B}}{T^4}\right)
  -\left(\frac{3}{\pi^2}\left[\frac{13}{6}T_{\mathfrak{q}}^B+2T_{\mathfrak{q}}^F\right]+\frac{\mathfrak{B}}{T^4}\right)H_{\mathfrak{q}}(V;T)\right).
\end{equation}
We note the Eq.(\ref{EDensity1}) has an important form expressing the energy density, the normalized energy density of the system $\varepsilon_{\mathfrak{q}}(V;T)/T^4$ is shown in Fig.(\ref{fig:FIG2}).
Traditionally, the quantity $\varepsilon_{\mathfrak{q}}(V;T)/T^4$ has been led to the interpretation the number of effective degrees of freedom in the system. The liberation/melting of the partonic degrees of freedom locked in the hadronic phase, as shown in Fig.(\ref{fig:FIG2}) of the normalised energy density during the QCD deconfinement phase transition, is logically persuasive. In the thermodynamical limit, this deconfinement causes a finite smooth discontinuity in finite volume to become a finite sharp discontinuity, which is connected to the latent heat of the first-order deconfinement phase transition. Over virtually the entire temperature range, there is a strong size dependency. A step function, which changes into a $\delta$-function in the specific heat $c_T(V;T)$ and might serve as an indication for the finite volume transition point, can be used to quantitatively characterise the finite discontinuity.
\begin{figure}[h]
  \centering
 \includegraphics[width=400pt]{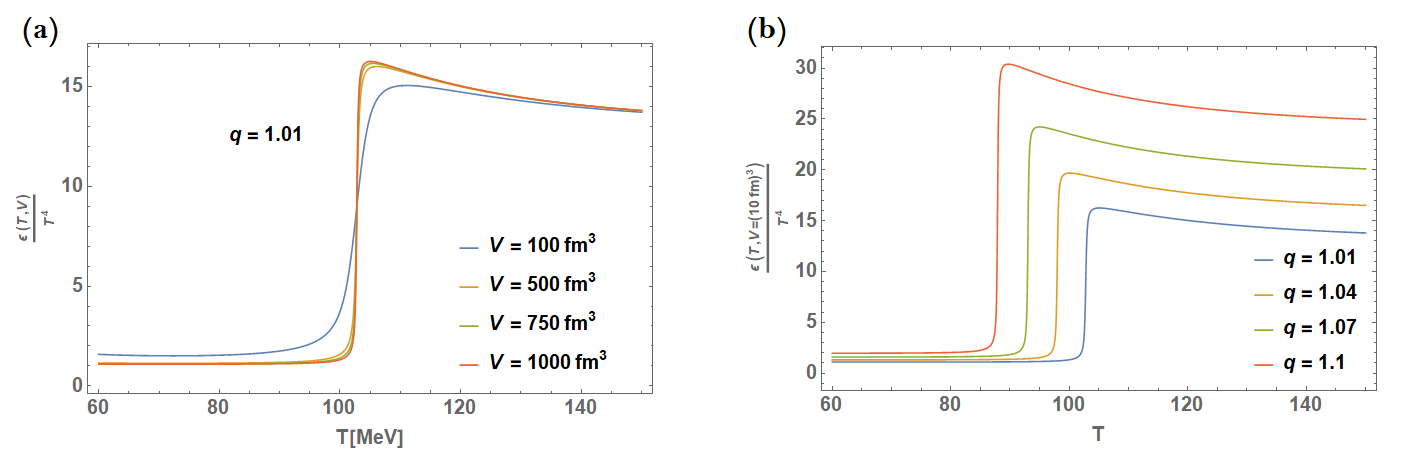}
  \caption{(a) The energy density versus temperature $T$, when $\mathfrak{q}=1.01$ with set volumes $V=(0.1,0.25,0.5,1)(10fm)^3$.(b) The energy density versus temperature $T$, when $V=(10fm)^3$ with set Tsallis parameter $\mathfrak{q}=(1.01,1.04,1.07,1.1)$.}\label{fige}
\end{figure}
In order to further investigate the nature of the phase transition, we have analysed certain density energy, as a function of the temperature. This is seen in Fig.(\ref{fige}). With the increase of the $\mathfrak{q}$-values, the energy density $\varepsilon_{\mathfrak{q}}/T^4$ grows faster, exhibiting a continuous abrupt increase in its value for particular temperatures, suggesting a probable shift to greater degrees of freedom. Then after reaching a maximum value for all $\mathfrak{q}$-values, it starts dropping at higher temperatures and moves to a saturation zone. This conduct can be understood as follows. As the temperature increases, more and more heavy resonances start contributing to the $\varepsilon_{\mathfrak{q}}/T^4$, and therefore one sees a strong increase in the $\varepsilon_{\mathfrak{q}}/T^4$. At infinite temperature, the system starts acting like a gas of massless particles, and $\varepsilon_{\mathfrak{q}}/T^4$ finally goes to a constant value. The fast increase with the parameter $\mathfrak{q}$ happens because the distribution deviates more and more from a Boltzmann distribution and the tails with large momentum contribute more and more as $\mathfrak{q}$ grows.

\begin{figure}[h]
\centerline{\includegraphics[width=250pt]{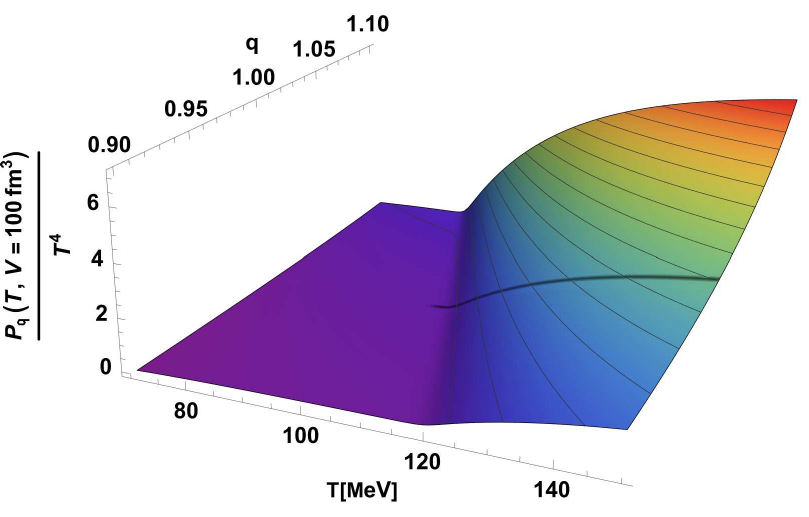}}
\caption{3-Dim normalized pressure $\frac{p_{\mathfrak{q}}(T)}{T^{4}}$ of the system as a function of $(T)$, when the volume $V=100fm^3$. \label{fig:FIG4}}
\end{figure}

\begin{figure}[h]
  \centering
 \includegraphics[width=400pt]{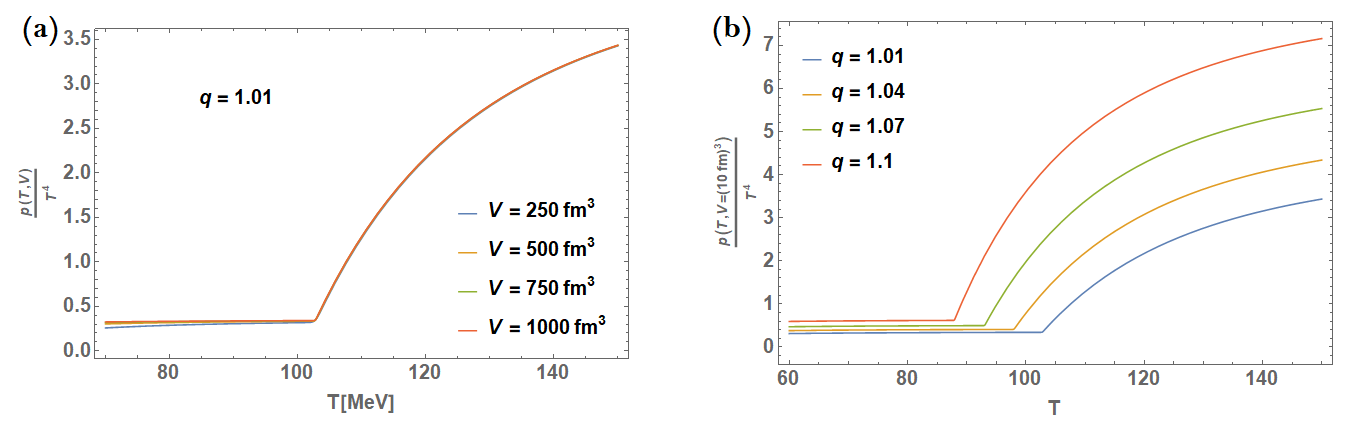}
  \caption{(a) The normalized pressure versus temperature $T$, when $q=1.01$ with set volumes $V=(0.25,0.5.0.75,1)(10fm)^3$.(b) The normalized pressure versus temperature $T$, when $V=(10fm)^3$ with set Tsallis parameter $\mathfrak{q}=(1.01,1.04,1.07,1.1)$.}\label{figp}
\end{figure}

\begin{figure}[h]
\centerline{\includegraphics[width=250pt]{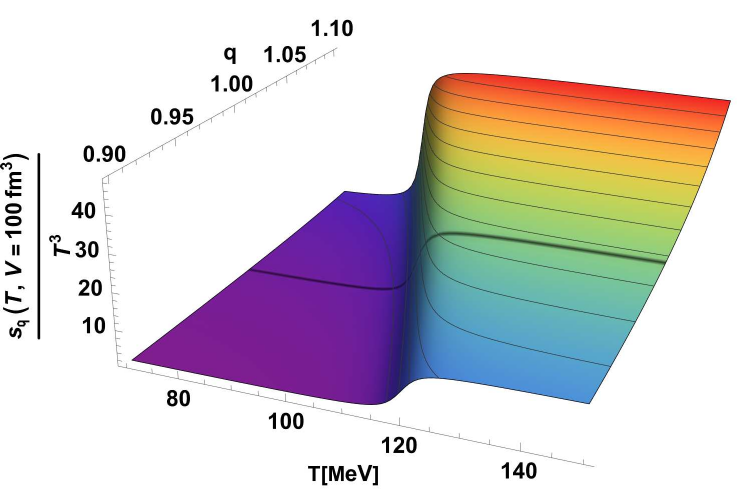}}
\caption{3-Dim normalized Entropy density $\frac{s_{\mathfrak{q}}(T)}{T^{3}}$ of the system as a function of $(T)$, when volume $V=100fm^3$. \label{fig:FIG7}}
\end{figure}

The definition of the pressure in thermodynamical is given by \cite{Tannoudji1997,Feynman1982,Balian2007}
\begin{equation}\label{Press}
p_{\mathfrak{q}}(V;T) =T \left\langle \left( \frac{\partial LnZ_{\mathfrak{q}}(h,V;T)}{\partial V}\right) \right\rangle,
\end{equation}
and after a simple procedure, we obtain the final formal of the total pressure, which show the contribution of the pressure during phase transition in our system,
\begin{equation}
\frac{p_{\mathfrak{q}}(V;T)}{T^4} =\left(\left(\frac{2}{\pi^2}\left[\frac{4}{3}T_{\mathfrak{q}}^B+2T_{\mathfrak{q}}^F\right]-\frac{\mathfrak{B}}{T^4}\right)
  -\left(\frac{1}{\pi^2}\left[\frac{13}{6}T_{\mathfrak{q}}^B+2T_{\mathfrak{q}}^F\right]-\frac{\mathfrak{B}}{T^4}\right)H_{\mathfrak{q}}(V;T)\right).
  \label{Pressure}
\end{equation}
The plot of the estimated pressure normalised by $T^4$ is shown in Fig.(\ref{fig:FIG4}). Just after the transition temperature, and especially when the volume approaches the thermodynamical limit, a fast increase in pressure can be noticed.
It is obvious that below the transition temperature, the pressure remains constant with the SB value in the HG phase, then it rapidly increases at $T_{0}(V)$ and continues to grow, albeit slowly. As a function of temperature, the pressure continues to grow in a monotonic manner. A behaviour that is compatible with the phenomena of phase coexistence.
The behaviour of $p_{\mathfrak{q}}/T^4$ in the Fig.(\ref{figp}), its response increases with temperature growth and different values of $\mathfrak{q}$ and gradually stabilizes.
The entropy density $s_{\mathfrak{q}}(V;T)$ may also be calculated using the same scheme from the conventional definition,
\begin{equation}\label{ENDensity}
  s_{\mathfrak{q}}(V;T) =\frac{1}{V} \left<\frac{\partial}{\partial T}\left( T Ln Z_{\mathfrak{q}}(h,V;T)\right) \right>.
\end{equation}
the entropy density is related to the order parameter, so after we use Eq.(\ref{ENDensity}), the entropy density will be written as
\begin{equation}\label{ENDensity1}
  \frac{s_{\mathfrak{q}}(V;T)}{T^3}=\frac{8}{\pi^2}\left(\left[\frac{4}{3}T_{\mathfrak{q}}^B+2T_{\mathfrak{q}}^F\right]
  -\left[\frac{13}{12}T_{\mathfrak{q}}^B+T_{\mathfrak{q}}^F\right]H_{\mathfrak{q}}(V;T)\right).
\end{equation}
The Heaviside step-function can use to parameterize the order parameter in the thermodynamical limit in a straightforward method as
$$\lim_{\substack{V \to \infty \\ \mathfrak{q} \to 1}}
           H_{\mathfrak{q}}(V;T)=1-\Theta(T-T_0(\infty)).$$
We should remark that this relation has an important form expressing the entropy density; the normalised entropy density of the system $s_{\mathfrak{q}}(V;T)/T^3$ is shown in Fig.(\ref{fig:FIG7}).

Traditionally, the quantity $s_{\mathfrak{q}}(V;T)/T^3$ has been interpreted as a measure of the number of effective degrees of freedom in the system. The liberation/melting of the partonic degrees of freedom locked in the hadronic phase, as seen in Fig.(\ref{fig:FIG7}) of the normalised entropy density in the QCD deconfinement phase transition, is logically persuasive. In the thermodynamical limit, this deconfinement causes a finite smooth discontinuity at finite size to become a finite sharp discontinuity, which is connected to the latent heat of the first-order deconfinement phase transition. Over virtually the entire temperature range, there is a strong size dependency.
\begin{figure}[h]
  \centering
 \includegraphics[width=400pt]{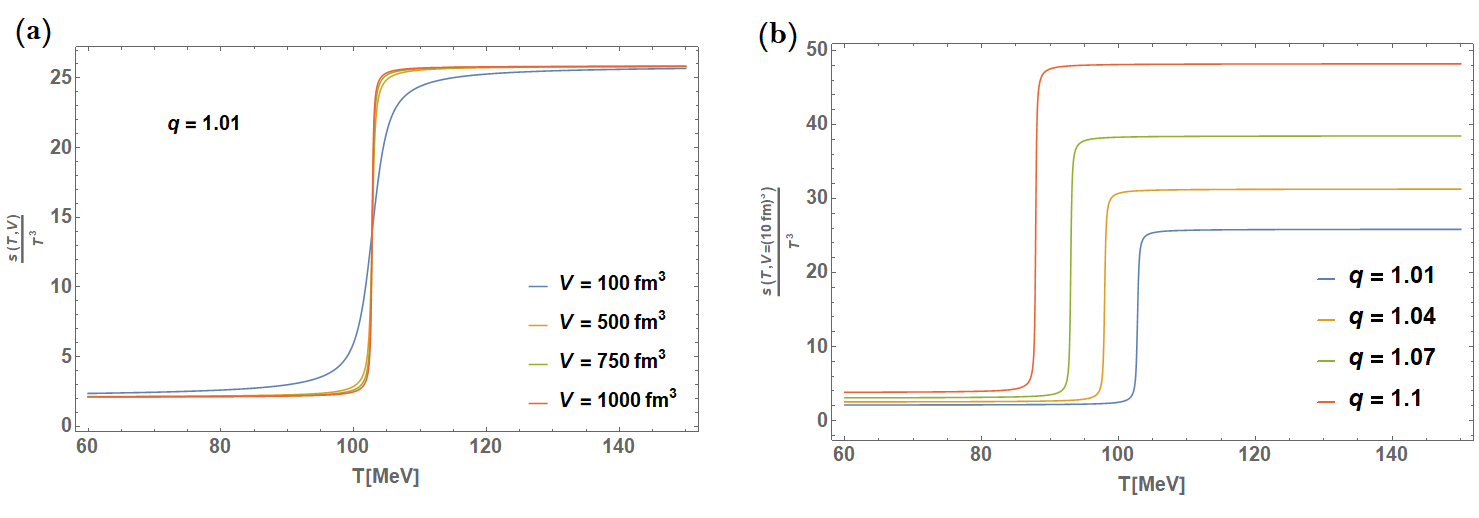}
  \caption{(a) The Entropy density versus temperature $T$, when $\mathfrak{q}=1.01$ with set volumes $V=(1,0.25,0.5,1)(10fm)^3$.(b) The Entropy density versus temperature $T$, when $V=(10fm)^3$ with set Tsallis parameter $\mathfrak{q}=(1.01,1.04,1.07,1.1)$.}\label{fign}
\end{figure}
When approaching the thermodynamical limit, at a transition temperature, the first-order character of the transition is demonstrated by the step-like rise or sharp discontinuity of each of the order parameter, as well as the normalised energy and entropy densities, which reflects the presence of latent heat accompanying the phase transition. The quantities $\varepsilon_{\mathfrak{q}}/T^4$ and $s_{\mathfrak{q}}/T^3$ are traditionally interpreted as a measure of the number of effective degrees of freedom \cite{Barter}. The temperature rise causes ``melting'' of the constituent degrees of freedom ``frozen'' in the hadronic case, causing the energy and entropy densities to reach their plasma values. Because of the significant thermodynamical fluctuations, the probability of the presence of the PP phase below the critical point, and of the pionic phase above the critical point are limited in small systems, the phase transition is rounded.

Moreover, we note from Figs.(\ref{fige}), (\ref{figp}) and (\ref{fign}),  when $\mathfrak{q}$ goes to unit, $p_{\mathfrak{q}}/T^4$, $\varepsilon_{\mathfrak{q}}/T^4$, and $s_{\mathfrak{q}}/T^3$, all tend to their SB limit. However, as $\mathfrak{q}$ increases, they increase rapidly until they exceed their corresponding SB limits. Taking $\varepsilon_{\mathfrak{q}}/T^4$ as an example, for the QCD deconfinement phase transition from HG to PP and the temperature is fixed at 0.15 GeV. When $\mathfrak{q}=1$, the value of $\varepsilon_{\mathfrak{q}}/T^4$ is 13.036, very close to the SB limit 12.18. But when $\mathfrak{q}=1.1$, the value of $\varepsilon_{\mathfrak{q}}/T^4$ is 24.973, which is increased by 91\%.

Furthermore, from Figs.(\ref{fige}) and (\ref{fign}), we find that the response patterns of the density energy and entropy for $\mathfrak{q}$ are almost the same. They reach a maximum near the transition point $T_0$, then decrease and tend to be stable. The relationship becomes more straightforward when the pressure is expressed as a function of the energy density:
 \begin{eqnarray}\label{pressureandDE}
   \theta_{\mathfrak{q}}(T;V) &=& \epsilon_{\mathfrak{q}}(T;V)-3p_{\mathfrak{q}}(V;T), \\
 \nonumber &=& 4\mathfrak{B}\left(1-H_{\mathfrak{q}}(V;T)\right).
 \end{eqnarray}
\begin{figure}[h]
\centerline{\includegraphics[width=250pt]{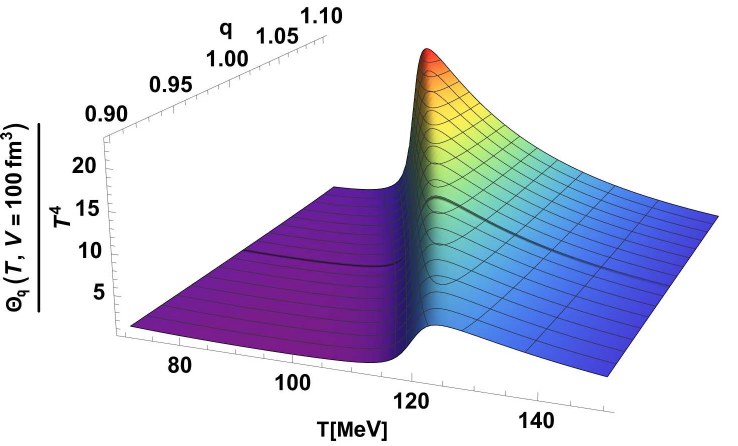}}
\caption{3-Dim trace anomaly $\frac{\theta_{\mathfrak{q}}(T)}{T^{4}}$ of the system as a function of $(T,V)$, when the volume $V=100fm^3$. \label{fig:FIG5}}
\end{figure}
The interaction measure is the difference between $\varepsilon_{\mathfrak{q}}(V;T)$ and $3p_{\mathfrak{q}}(V;T)$ and also known as the trace anomaly $\theta_{\mathfrak{q}}(V;T)$, this quantity is vanishing in massless gas and non-interacting, whereas appears in the case of an interacting system. The relationship between the non-zero value of the interaction measure and TS and the vacuum contribution has been thoroughly understood in our QCD MIT Bag Model. The expression $4\mathfrak{B}(1-H_{\mathfrak{q}}(V;T)$ appears in our EoS, in which TRF $(1-H_{\mathfrak{q}}(V;T))$. This term distinguishes itself from the variation with the relativistic EoS, commonly used: $3p_{\mathfrak{q}}(V;T)=\varepsilon_{\mathfrak{q}}(V;T)$. It is related to everything that occurs during the deconfinement phase transition, therefore its non-perturbative origin is obvious.

\begin{figure}[h]
  \centering
 \includegraphics[width=400pt]{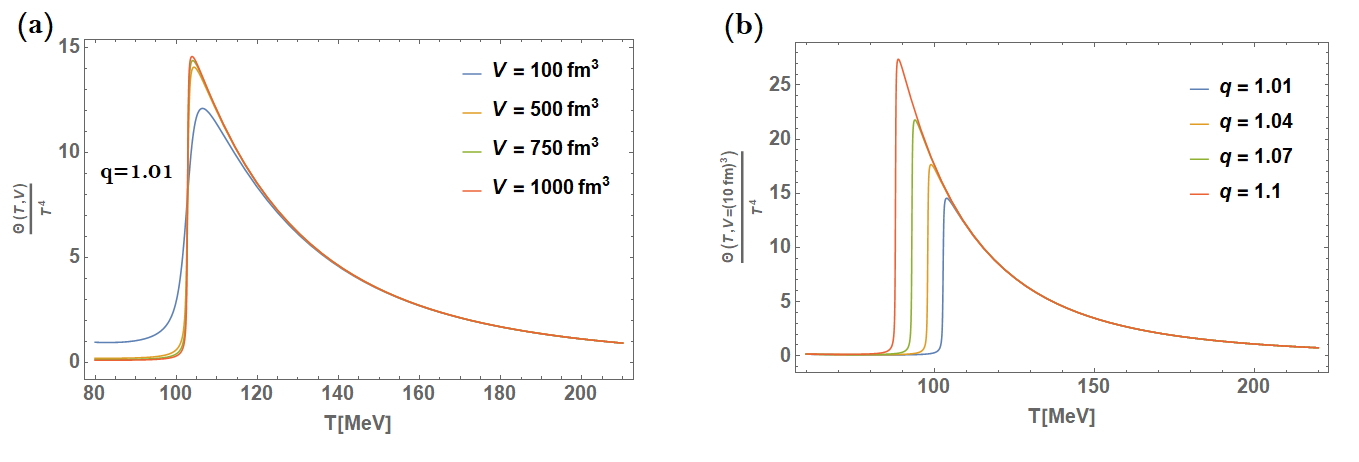}
  \caption{(a) The trace anomaly versus temperature $T$, when $\mathfrak{q}=1.01$ with set volumes $V=(0.1,0.5,0.75,1)(10fm)^3$.(b) The trace anomaly versus temperature $T$, when $V=(10fm)^3$ with set Tsallis parameter $\mathfrak{q}=(1.01,1.04,1.07,1.1)$.}\label{figi}
\end{figure}

The graph in Fig.(\ref{fig:FIG5}) shown the difference of the interaction measure $\theta_{\mathfrak{q}}(V;T)$ as a function of temperature for various values of the Tsallis parameter. Our findings appear to be in good convention with the interaction measure common behaviour. In particular, $\theta_{\mathfrak{q}}(V;T)$, climbs fast after the transition point with the values of Tsallis parameter, after that slowly drops at high temperatures, allowing to form a maximum point appear just beyond in the transition region this point changes with the value of the Tsallis parameter.
In both low and high temperatures, the Tsallis parameter dependency is small, as seen in Fig.(\ref{figi}). However, in the middle area, about the maximum, this dependency is apparent. The system is not in an optimal condition just above the Tsallis parameter transition temperature. The mutual interactions between partons result in a non-vanishing interaction measure, which agrees well with many models, such as hot lattice QCD computations. The fact that the pressure, energy density, and entropy density of PP are all far below ideal gas values even at high temperatures suggests that there are still significant interactions between the partons, showing the non-ideal character of the strongly connected PP.
The trace anomaly is related to $p_{\mathfrak{q}}/\varepsilon_{\mathfrak{q}}$, this quantity help to measures the deviation from EoS of an ideal gas $\varepsilon_{\mathfrak{q}}=3p_{\mathfrak{q}}$ due to interactions and/or finite quark masses.

\begin{figure}[h]
\centerline{\includegraphics[width=250pt]{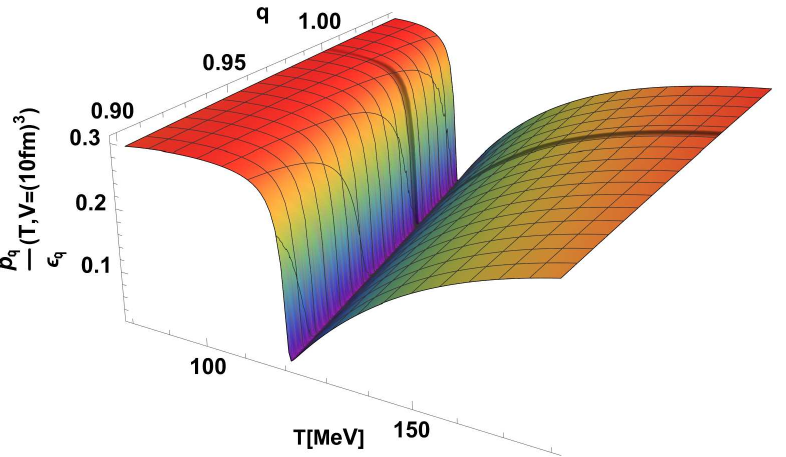}}
\caption{3-Dim Plot of $\frac{p_{\mathfrak{q}}}{\varepsilon_{\mathfrak{q}}}(T;V)$ of the system as a function of $(V;T)$, when the volume $V=(10fm)^3$. \label{fig:FIG6}}
\end{figure}

It has been observed that HG, $\frac{p_{\mathfrak{q}}}{\varepsilon_{\mathfrak{q}}}(V;T)$, has a minimum at the phase transition point, as shown in Fig.(\ref{fig:FIG6}). In other words, at $T_{0}(V)$, the EoS in the QCD MIT-bag model becomes softer, and the system develops slower at the $p_{\mathfrak{q}}(V;T)$/$\varepsilon_{\mathfrak{q}}(V;T)$ ratio's minimum. This minimum refers to as the softest point of the EoS.

The ratio $\frac{p_{\mathfrak{q}}}{\varepsilon_{\mathfrak{q}}}(V;T)$ versus $\varepsilon_{\mathfrak{q}}(V;T)$ demonstrates little dependence on volume in the low temperatures as well as the high temperatures. With increasing energy density at finite volumes, the ratio $p_{\mathfrak{q}}/\varepsilon_{\mathfrak{q}}$ continues to drop until it reaches the softest point. Beyond this point, the ratio reverses the direction of its variation and increases with the energy density. It is said that the low pressure at this point inhibits the system from rapidly expanding and cooling, resulting in a slow-burning regime. The minimum of the ratio grows as the volume decreases, and the softening of the EoS becomes less evident. The pressure remains constant throughout various energy densities, indicating that the system does not do mechanical work. This region is pointed out as the softest area of the EoS, where a distinct finite volume effect can be seen. We also observe that for large volumes, the ratio's initial value in the pure HG phase is identical to the asymptotic value, which has been achieved well beyond the softest point in the pure phase of PP.
When a deconfinement phase transition occurs from pionic gas to PP, the pressure change is much less, resulting in a modest sound velocity, which is determined by the pressure gradient with regard to energy density. The collective dynamics of the hot and dense matter produced in URHIC are predicted to be significantly influenced by this soft area in our system's EoS. A low sound velocity, in particular, slows the expansion of compressed matter and reduces the transverse collective flow. As in the hydrodynamics model, the softening EoS results in a minimum in the incident energy dependence of the transverse collective flow and a delayed expansion of the compressed matter.
The softest point of the EoS is found to be $\left(\frac{p_{\mathfrak{q}}}{\varepsilon_{\mathfrak{q}}}\right)_{min}\simeq0.03$ for each volume, and this is in agreement with what it was found in lattice QCD with colour \cite{Ejiri2006,Cheng2008}.
\begin{figure}[h]
  \centering
 \includegraphics[width=400pt]{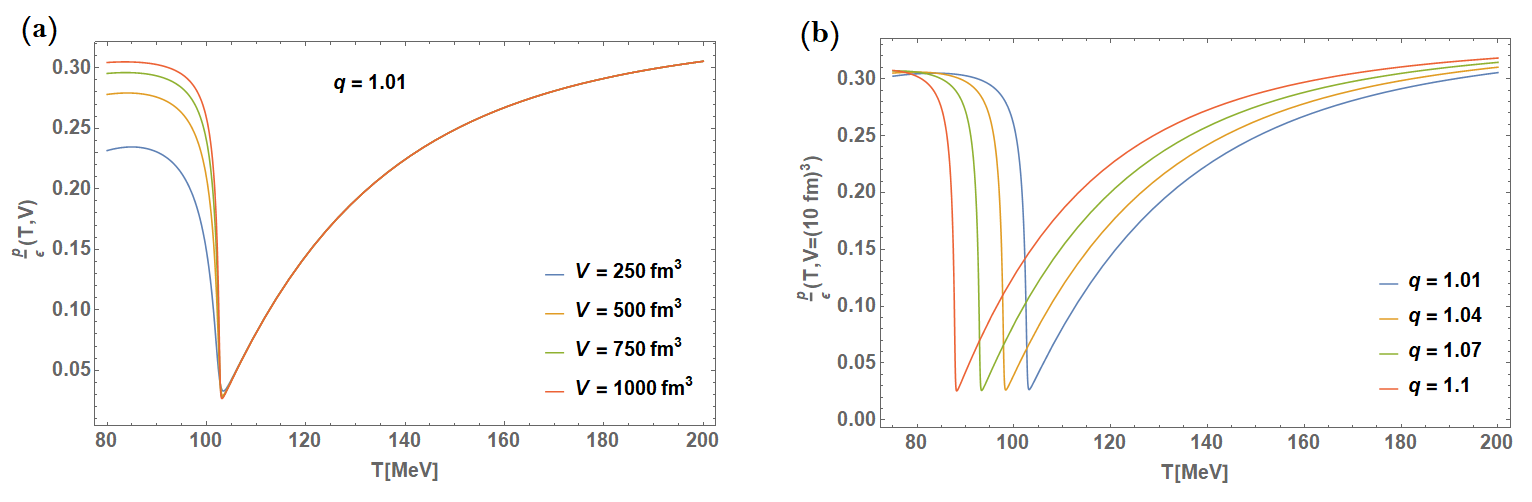}
  \caption{(a) Plot of $\frac{p_{\mathfrak{q}}}{\varepsilon_{\mathfrak{q}}}(T)$  versus temperature $T$, when $\mathfrak{q}=1.01$ with set volumes $V=(0.25,0.5,0.75,1)(10fm)^3$.(b) The Plot of $\frac{p_{\mathfrak{q}}}{\varepsilon_{\mathfrak{q}}}(V;T)$ versus temperature $T$, when $V=(10fm)^3$ with set Tsallis parameter $\mathfrak{q}=(1.01,1.04,1.07,1.1)$.}\label{figr}
\end{figure}
The Fig.(\ref{figr}) shows the behaviour of $p_{\mathfrak{q}}/\varepsilon_{\mathfrak{q}}$ with different values of $\mathfrak{q}$, near the transition point $T_0$ it has a dip and then approaches the ideal gas value of $1/3$ at high enough temperatures.
Now, we can use Eq.(\ref{pressureandDE}) to rewrite
\begin{equation}\label{ch8tn}
  \frac{\theta_{\mathfrak{q}}}{\varepsilon_{\mathfrak{q}}}(V;T)=1-3\frac{p_{\mathfrak{q}}}{\varepsilon_{\mathfrak{q}}}(V;T),
\end{equation}
and
\begin{equation}\label{ch8tn1}
  \frac{\varepsilon_{\mathfrak{q}}}{\theta_{\mathfrak{q}}}(V;T)\approx1+3\frac{p_{\mathfrak{q}}}{\varepsilon_{\mathfrak{q}}}(V;T),
\end{equation}
this approximation came from \cite{Ghosh2006,Schaefer2010}, we note $p_{\mathfrak{q}}/\varepsilon_{\mathfrak{q}}$ is in good agreement at both low temperature and high temperature around the transition region.

At the intermediate temperature between the low and high temperatures from the transition point, the speed of sound is slightly larger than $p_{\mathfrak{q}}/\varepsilon_{\mathfrak{q}}$. Therefore, it can be known from Eq.(\ref{ch8tn}) that near the transition point $T_0$, the minimum value of $p_{\mathfrak{q}}/\varepsilon_{\mathfrak{q}}$ will cause the maximum value of $\theta_{\mathfrak{q}}/T^4$. At the high temperature limit, $p_{\mathfrak{q}}/\varepsilon_{\mathfrak{q}}$ tends to $1/3$ and $\theta_{\mathfrak{q}}/T^4$ tends to zero. From Fig.(\ref{figi}), we can clearly see that it has a peak near the transition point $T_0$ and then tends to zero.
\begin{figure}[h]
  \centering
 \includegraphics[width=200pt]{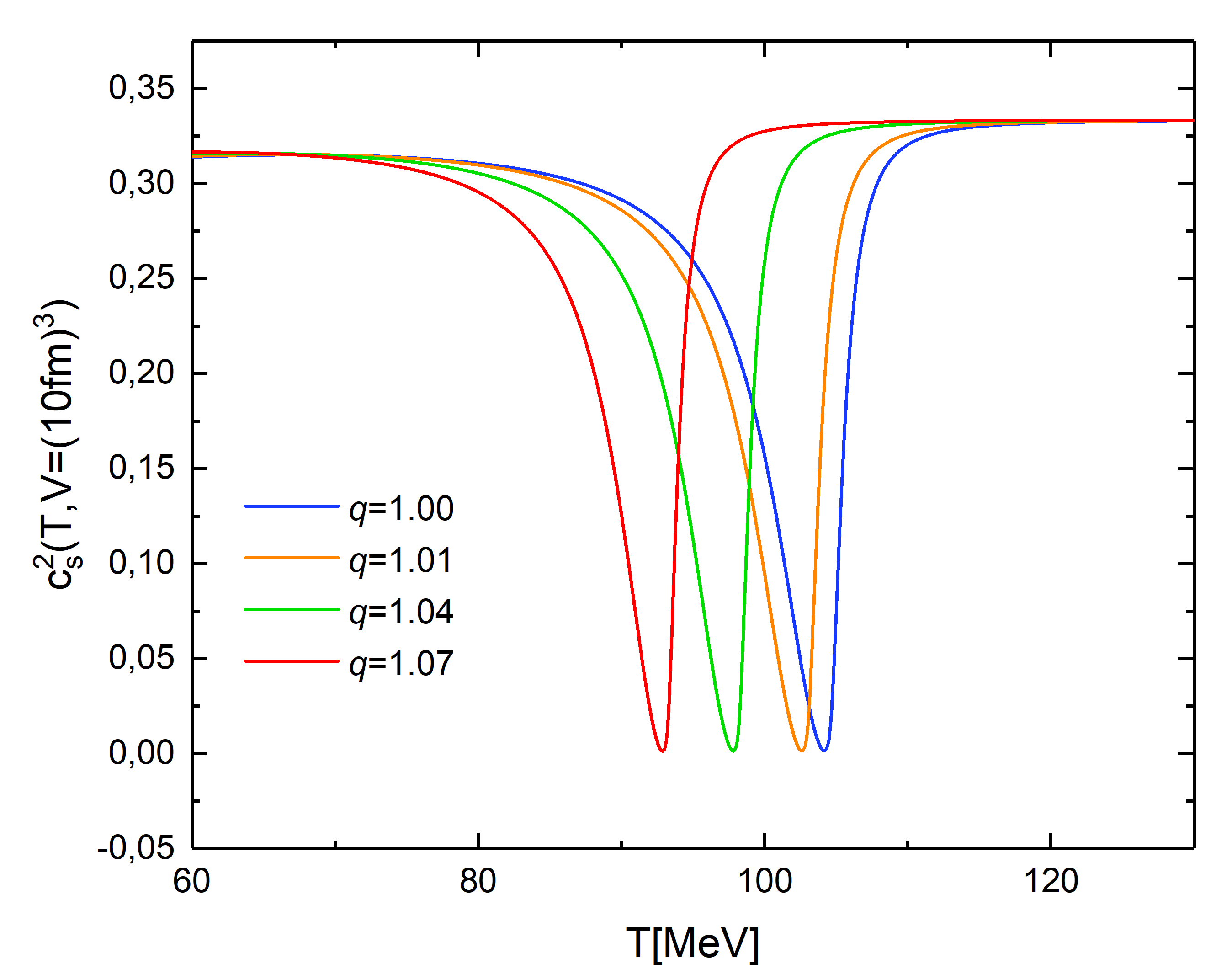}
  \caption{Sound velocity $c_{s}^{2} (V=(10fm)^3;T )$ vs temperature for different values of $\mathfrak{q}$.}\label{figs}
\end{figure}
The velocity of sound squared is plotted in Fig.(\ref{figs}) as a function of temperature, for different $\mathfrak{q}$ values, and it is obvious from the curves that for temperatures well below and/or above a transition temperature, it approaches the value for an ultra-relativistic ideal gas $c_s^2=1/3$.
In a finite volume, $c_s^2$ decreases in the transition region, and has a minimum at the effective transition temperature $T_0(V)$ as displayed in Figure. \ref{figs}. Near the $T_0$ transition temperature for $\mathfrak{q}=1$. At high enough temperatures, the speed of sound dips and then approaches the ideal gas value of $1/3$. Because the finite-size effect is a part of the nonextensive effect.

\section{\textbf{Conclusion}}
The work has clarified the role of the finite volume of the system and the non-extensive Tsaills distributions on the behaviour of some response functions in the vicinity of the transition point during the deconfinement phase transition from HG to PP. The non-extensive Tsaills distributions lead to shifting of the transition point was observed at finite volume. The sharp transition which is observed in the thermodynamical limit in some thermodynamical quantities like the order parameter, energy density, pressure, entropy density and sound velocity at transition temperature $T_0(\infty)$, is rounded in finite volumes, and the variations of those thermodynamical quantities are perfectly smoothed on the limited range of temperature. This done, without the colorlessness condition, the finite volume effect on the position of the transition point (shifting) is observable. When the colorlessness condition is taken into account, a shift in the transition point is observed for different system volumes. In contrast, without this condition, the finite-volume effect on the position of the transition point becomes negligible \cite{Mhamed2015,Rozynek2009}.
We addressed the effect of the parameter $\mathfrak{q}$ on the deconfinement phase transition, as well as different thermodynamical characteristics at finite temperature without the chemical potential. We discovered that the SB limit is connected to the statistics used. In the TS, for example, the thermodynamical quantities $\varepsilon_{\mathfrak{q}}/T^4$, $p_{\mathfrak{q}}/T^4$, and $s_{\mathfrak{q}}/T^4$ all increase with $\mathfrak{q}$, surpass their normal SB limits, and trend to a new $\mathfrak{q}$-related Tsallis limit at high enough temperatures. However, owing to an unexpected cancellation, the high temperature limit of $p_{\mathfrak{q}}/\varepsilon_{\mathfrak{q}}$ is still its SB limit of $1/3$. In this work, the results show that the transition point varies with the nonextensivity parameter $\mathfrak{q}$, but remains constant at a specific value of $\mathfrak{q}$, even for different volumes \cite{Mhamed2018,Rozynek2009}.
Furthermore, we found some analogous between the nonextensive effect and the finite-size effect. Furthermore, the results show that with increasing values of the Tsallis nonextensivity parameter $\mathfrak{q}$, the transition temperature decreases and the transition becomes sharper \cite{Tsallis1988}. This behavior implies that stronger nonextensive effects (larger $\mathfrak{q}$) enhance deconfinement, as fluctuations and long-range correlations facilitate the formation of the QGP \cite{Bhattacharyya2016,Lavagno2013,Cleymans2012a}. However, because to an unexpected cancellation, the parameter $\mathfrak{q}$ has no effect on the high temperature limit for $p_{\mathfrak{q}}/\varepsilon_{\mathfrak{q}}$ and $\theta_{\mathfrak{q}}/T^4$ \cite{Zhao2020}. We obtained that their responses to $\mathfrak{q}$ are various. TRF have the highest response to $\mathfrak{q}$ near the transition point $T_0(V)$.
Future investigations will extend this study within TS by incorporating the colorlessness condition and, additionally, considering the effects of the chemical potential \cite{Zhang2025}.

\end{document}